\documentclass[showpacs,preprintnumbers,amsmath,amssymb,twocolumn,superscriptaddress,nofootinbib,nobibnotes,longbibliography]{revtex4-2}
\usepackage{color}
\usepackage[unicode=true,colorlinks=true,citecolor=blue]{hyperref}
\usepackage{epstopdf}
\usepackage{graphicx}
\usepackage[normalem]{ulem}
\usepackage{tabularx}

\usepackage{dcolumn}% Align table columns on decimal point
\usepackage{multirow}
\usepackage{bm}% bold math
\usepackage{upgreek}
\usepackage{physics}
\DeclareMathOperator\Arctanh{Arctanh}

\usepackage[T2A]{fontenc}
\usepackage[cp1251]{inputenc}

\begin{document}

\title{
Electrical Magnetochiral current in Tellurium
}

\author{L.~E.~Golub} 
\affiliation{Terahertz Center, University of Regensburg, 93040 Regensburg, Germany}	
\author{E.~L.~Ivchenko} 	
\affiliation{Ioffe Institute,  	194021 St.~Petersburg, Russia}
\author{B.~Spivak} 	
\affiliation{University of Washington, Seattle, WA 98195, USA}

%\date{\today}

\begin{abstract}
We have studied theoretically the effect of Electrical Magneto-Chiral Anisotropy (eMChA) in $p$-type tellurium crystals.  It is shown that the terms $k_i B_j$ in the hole Hamiltonian, linear both in the wave vector ${\bm k}$ and the magnetic field ${\bm B}$, do not lead to the eMChA and one needs to include the higher-order terms like $k_i^3 B_j$. Two microscopic mechanisms of the effect are considered. In the first one only elastic scattering of holes by impurities or imperfections are taken into consideration only. In the second mechanism, besides the elastic scattering processes the hole gas heating and its energy relaxation are taken into account. It is demonstrated that he both contributions to the magneto-induced rectification are comparable in magnitude. The calculation is performed by using two independent approaches, namely, in the time relaxation approximation and in the limit of of small chiral band parameter $\beta$. A bridge is thrown between the eMChA and magneto-induced photogalvanic effects.
\end{abstract}

\maketitle

\section{Introduction}
Tellurium is an elemental chiral crystal with a D$_3$ point symmetry. It has a natural optical activity~\cite{Nomura,Dubinskaya}, and it is tellurium where the Circular Photogalvanic effect~\cite{ExpTe,Pressure}, electric-current induced optical activity~\cite{Farbshtein,sha12} and bulk Circular Photon Drag effect~\cite{CPDE_Te} were discovered; Sakano et al. has for the first time verified experimentally the spin texture of the right- and left-handed tellurium by the ARPES and SARPES  measurements~\cite{spin_textures}. Recently, Rikken and Avarvari observed the effect of Electrical Magneto-Chiral Anisotropy (eMChA) in Te crystals~\cite{Rikken2019}. This effect manifests itself as an additional contribution to the sample resistance $R=R_0(1 + \gamma B I)$, where $R_0$ is a constant, $B$ is the magnetic field strength,  $I$ is the electric current, and the coefficient of bilinear magneto-electric resistance $\gamma$ describes a rectification by the sample, see Refs.~\cite{Rikken2021,Nagaosa} for reviews.
Earlier, the effect of chirality (or non-reciprocity) in magnetotransport has been observed in a number of other gyrotropic materials: distorted bismuth wires~\cite{Rikken2001}, carbon nanotubes~\cite{Rikken2002,CN2005}, crystals of chiral salt (DM-EDT-TTF)$_2$ClO$_4$~\cite{Salts}, polar semiconductor crystal BiTeBr~\cite{BiTeBr}, topological insulators~\cite{TIa,He_2018,Dyrdal,Loss,eMCha_HgTe_Fert},  semimetals ZrTe$_5$~\cite{Ando}, WTe$_2$~\cite{WTe2} and $\alpha$-Sn~\cite{aSn}, and on the surface of SrTiO$_3$(111)~\cite{SrTiO3}. 

Theoretically, the eMChA effect has been considered for carbon nanotubes~\cite{Spivak,Spivak1}, Weyl semimetals of TaAs type~\cite{Nagaosa2016a}, semimetal ZrTe$_5$~\cite{Ando}~(Supplemental Material), surface states in topological insulators~\cite{Vignale} and molecular conductors~\cite{Tinvariant}. In the works~\cite{Vignale,Ando}, a calculation of the correction to the electric current $\delta j \propto E^2 B$, proportional to the squared electric field strength $E$ and linear in the magnetic field $B$, has been performed in the simplest approximation of a general relaxation time ($\tau$-approximation). This approach does not take into account a difference between quasimomentum and energy relaxations, or between elastic and inelastic relaxation processes of free charge carriers. In this paper we show that, with account for this difference, there are two independent microscopic mechanisms of eMChA. In a simplified form, the presence of two mechanisms can be explained as follows: Let us divide a correction to the charge carrier distribution function $\delta f_{\bm k} \propto E^2$  in two terms, $\delta f(\varepsilon_{\bm k})$ and $\delta f^{as}_{\bm k}$, where the first function depends on the carrier energy $\varepsilon_{\bm k}$ (${\bm k}$ is a wavevector), and the second function, $\delta f^{as}_{\bm k}$, is an asymmetric correction with zero average over the directions of the wavevector ${\bm k}$ at constant energy. The correction  $\delta f^{as}_{\bm k}$ is controlled by the momentum relaxation time $\tau_p$, while in order to calculate $\delta f(\varepsilon_{\bm k})$ one must account for inelastic processes of carrier-phonon interaction and, hence, introduce the energy relaxation time $\tau_{\varepsilon}$ which can be much longer than $\tau_p$. As noticed in Ref.~\cite{Spivak}, although the correction $\delta f(\varepsilon_{\bm k}) \propto \tau_\varepsilon$ by itself does not result in the electric current, its relaxation through interaction with phonons produces an asymmetric distribution of carriers in the $\bm k$ space with an extra multiplier $\tau_p/\tau_{\varepsilon}$. As a result, the mechanisms related to $\delta f^{as}_{\bm k}$ and $\delta f(\varepsilon_{\bm k})$ lead to comparable contributions to the electrical magneto-chiral current $\delta j \propto \tau_p^2 E^2 B$.

Here we consider both mechanisms resulting in eMChA of holes in the Te valence band. 
The paper is organized as follows. In Sec.~\ref{Macro}, macroscopic equations are presented. General consideration of eMChA effect in Te is given in Sec.~\ref{General}. In Sec.~\ref{tau_approx}, the eMChA current is estimated in the relaxation-time approximation. Sections~\ref{Elastic} and~\ref{Inelastic} are devoted to rigorous calculations of the contributions caused by the elastic and inelastic relaxation processes, respectively. The perturbative results in the lowest order in the chirality parameter are presented in Sec.~\ref{small_beta}. In Sec.~\ref{Disc_concl}, discussion of results is given, and Sec.~\ref{Summary} summarizes the paper.

\section{Macroscopic equations}
\label{Macro}
The phenomenon under study is described by a fourth-rank tensor in the expansion of the electric current density in powers of the electric field strength ${\bm E}$ and magnetic field ${\bm B}$
\begin{equation} \label{ji}
j_i = \sigma_{ij} E_j + \sigma^{(H)}_{ijk} E_j B_k +  G_{ijkl} E_j E_k B_l \:.
\end{equation}
The first two terms are allowed by any point symmetry, ${\bm \sigma}$ is the tensor of linear conductivity and $ \sigma^{(H)}_{ijk}$ is the Hall conductivity tensor. The eMChA effect  is represented by the magnetochiral tensor ${\bm G}$ symmetrical in indices $j$ and $k$. It is related by   
\[
G_{ijkl} \propto \gamma_{ij'k'l} \sigma_{j'j} \sigma_{k'k}
\]
with the tensor ${\bm \gamma}$ which is introduced in Eq.~(1) in Ref.~\cite{Rikken2019} and describes the second-harmonics generation
\[
E^{2 \omega}_i = \gamma_{ijkl} j^{\omega}_j j^{\omega}_k B_l \:,
\]
under conditions where the modulation period $T = 2 \pi/\omega$ exceeds by far all the microscopic times of the system.

In crystals of D$_3$ symmetry there are ten linearly-independent components of the $G_{ijkl}$ tensor with indices $zzzz$, $xxxx$, $zzxx$, $xxzz$, $zxxz$, $xzzx$, $xyyx$, $zxxy$, $xxyz$ and $xzxy$~\cite{Book}. Note that,  in this point group, the component $G_{xxyy}$ equals to $(G_{xxxx} - G_{xyyx})/2$. In Ref.~\cite{Rikken2019}, the following estimates are given: $12 G_{zzzx} \approx G_{xxxy} \approx 3 G_{xxxx}$, and the inequality $G_{zzzz} \ll G_{zzzx}$ is presented. This contradicts the point symmetry D$_3$ where the nonzero components $G_{zzzx}$ and $G_{xxxy}$ are forbidden.  In our work the attention is focused on the components $G_{zzzz}$ and $G_{xxxx}$ allowed by the symmetry, i.e., on the geometries ${\bm j} \parallel {\bm E} \parallel {\bm B} \parallel z$ (shortly $z$-eMChA geometry) and ${\bm j} \parallel {\bm E} \parallel {\bm B} \parallel x$ ($x$-eMChA). In these cases the Hall effect does not appear and hence is not discussed here. We use the notation $\delta {\bm j}$ for the electric magnetochiral (eMCh) current, or the third term in Eq.~ (\ref{ji}). The paper is devoted to consideration of this particular current.  

With acoount for the symmetry, the macroscopic relation between the correction to the current $\delta {\bm j}$ and the electric and magnetic vectors can be written in the following convenient form
\begin{widetext}
\begin{eqnarray} \label{Macro}
\delta j _z  &=& G^{(1)} E_z^2 B_z + G^{(2)} (E_x^2 + E_y^2) B_z + G^{(3)} [ (E_x^2 - E_y^2) B_y + 2 E_x E_y B_x ] + G^{(4)} E_z (E_x B_x + E_y B_y) \:, \\  \delta j _x  &=& G^{(5)} (E_x^2 + E_y^2) B_x + G^{(6)} E_z^2 B_x + G^{(7)} [ (E_x^2 - E_y^2) B_x + 2 E_x E_y B_y ] + G^{(8)} 2 E_xE_y B_z  \nonumber\\ && +\ G^{(9)} E_xE_zB_z + G^{(10)} E_z(E_x B_y + E_y B_x)\:,  \nonumber \\
\delta j _y  &=& G^{(5)} (E_x^2 + E_y^2) B_y +  G^{(6)} E_z^2 B_y + G^{(7)} [ -(E_x^2 - E_y^2) B_y + 2 E_x E_y B_x ] + G^{(8)} (E_x^2 - E_y^2) B_z  \nonumber \\ && +\ G^{(9)} E_yE_zB_z + G^{(10)} E_z(E_x B_x - E_y B_y)\:,  \nonumber
\end{eqnarray}
\end{widetext}
where $G^{(n)}~(n = 1\ldots10)$ are macroscopic parameters. The material relation between $\delta j_x$, $\delta j_y$ and the transverse components of vectors ${\bm E}$ and ${\bm B}$ has an axial symmetry and preserves its form at any orientation of the $x,y$ axes relative to the second-order symmetry axes C$_2$. 

\section{General consideration} \label{General}
The current $\delta j_z \propto G_{zzzz}$ is induced in the magnetic field ${\bm B} \parallel z$. In presence of this field,  the effective 2$\times$2 valence-band Hamiltonian in Te has the following form~\cite{Pressure,Bresler1970,Japan1970}
\begin{equation} \label{Hamilt}
\mathcal H = \mathcal A_1 k_z^2 + \mathcal A_2 k_\perp^2 + (\beta k_z + gB_z)\sigma_z + \Delta_2\sigma_x\:.
\end{equation}
Here $\bm k$ is a wavevector, $k_\perp^2 = k_x^2 + k_y^2$, $\sigma_x$ and $\sigma_z$ are the pseudospin Pauli matrices in the basis $\pm 3/2$ (the reducible representation ${\cal D} = H_4 + H_5$), $\Delta_2$ is the spin-orbit half-splitting of the valence-band states $$(|3/2\rangle \pm |-3/2\rangle)/\sqrt{2}$$ at the $H$ point of the Brillouin zone, the parameter $g$ describes the Zeeman effect, the parameters $\mathcal A_{1}, \mathcal A_{2}$ are responsible for parabolic scalar terms, and the coefficient $\beta$ determines strength of $k_z$-linear term, it has opposite signs in the two Te enantiomorphs D$_3^4$ and D$_3^6$ (or P3$_1$21 and P3$_2$21). Hereafter we use the hole representation and take ${\mathcal A_{1,2}>0}$.

\begin{figure}
	\centering \includegraphics[width=0.8\linewidth]{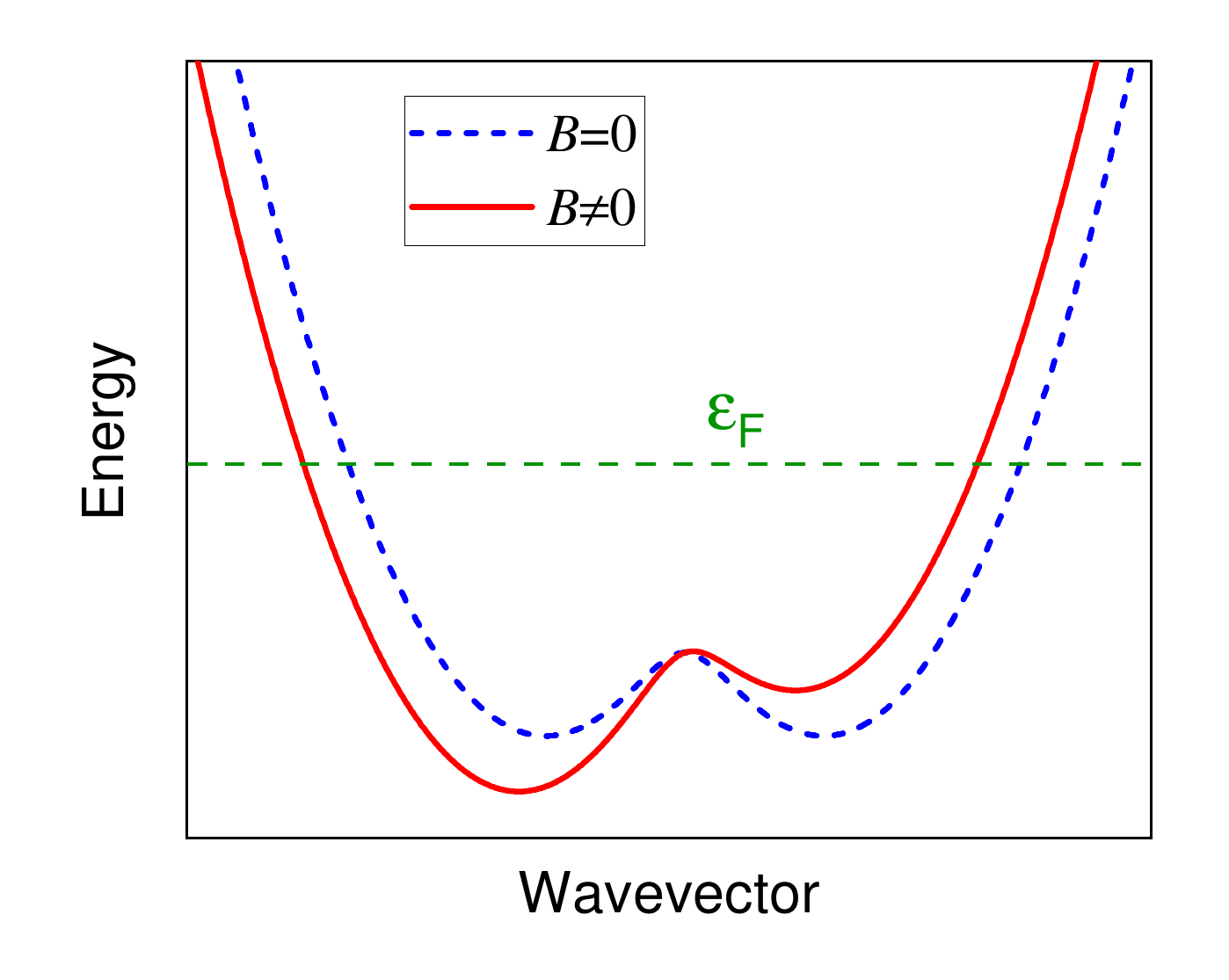}
	\caption{ 
	The lowest valence subband of tellurium in the hole
representation in the vicinity of the $H$ point. The curves show
the hole energy dispersion $\varepsilon_{\bm k}$ vs. $k_z$ at $k_\perp = 0$ in the absence
(dashed) and presence (solid) of the magnetic field $\bm B \parallel z$.}
	\label{Fig_spectrum}
\end{figure}

We study magnetoelectric transport of holes occupying the lowest valence band of Te (uppermost in the electron representation). According to Eq.~(\ref{Hamilt}) its energy dispersion relation is given by 
\begin{equation} \label{vv0dv}
\varepsilon_{\bm k} = \mathcal A_1 k_z^2 + \mathcal A_2 k_\perp^2 - \sqrt{\Delta_2^2+ (\beta k_z + g B_z)^2} + \Delta_2.
\end{equation}
Since we are interested in linear-${\bm B}$ effects, we make an expansion $\varepsilon_{\bm k} \approx \varepsilon_{\bm k}^0 + \delta\varepsilon_{\bm k}$,
where the zero-field energy is
\begin{equation}
\label{e0}
\varepsilon_{\bm k}^0 = \mathcal A_1 k_z^2 + \mathcal A_2 k_\perp^2 - \sqrt{\Delta_2^2+\beta^2 k_z^2} + \Delta_2\:,
\end{equation} 
and the correction
\begin{equation}
\label{d_E_k}
\delta\varepsilon_{\bm k} = - g B_z\eta(k_z), \qquad 
\eta={\beta k_z\over \sqrt{\Delta_2^2+\beta^2 k_z^2}}\:.
\end{equation}
The hole energy dispersion at zero magnetic field and at $B_z \neq 0$ is illustrated in Fig.~\ref{Fig_spectrum}.

At $B_z=0$ the eigenvectors of the Hamiltonian~(\ref{Hamilt}) are two-component columns
\begin{equation}
\label{u_eta}
u_{k_z}^0 = \frac{1}{\sqrt{2}} \left[ \begin{array}{c} \sqrt{1+\eta(k_z)}  \\
 \sqrt{1-\eta(k_z)} \end{array}  \right]\: .
\end{equation}
The dispersion $\varepsilon_{\bm k}^0$ has the camel's back shape with the energy minimum $\varepsilon_m = - \Delta^2 {\cal A}_1/\beta^2$. At fixed hole energy $\varepsilon_{\bm k}^0 = \varepsilon \geq \varepsilon_m$  the values of $k_{\perp}^2$ lie in the range between 0 and $\sqrt{(\varepsilon - \varepsilon_m)/{\cal A}_2}$ while the values of $k_z$ fill the the range $K_z(\varepsilon)$ containing two intervals $[- \kappa(\varepsilon),- \kappa'(\varepsilon)]$ and $[\kappa'(\varepsilon), \kappa(\varepsilon] $, where
\begin{eqnarray}
\label{kappa_def}
&&\kappa(\varepsilon)= \sqrt{\varepsilon -\Delta + \sqrt{\Delta^2 + \beta^2\varepsilon/\mathcal A_1}\over \mathcal A_1}\:, \\ && \kappa'(\varepsilon)= \sqrt{\varepsilon -\Delta - \sqrt{\Delta^2 + \beta^2\varepsilon/\mathcal A_1}\over \mathcal A_1}\:,\nonumber
\end{eqnarray}
and $\Delta = \Delta_2 - \beta^2/(2\mathcal A_1)$. For $\varepsilon > 0$, the value of $\kappa'$ should be set to 0 and the range $K_z(\varepsilon) = [- \kappa(\varepsilon), \kappa(\varepsilon)]$.

For calculation of the $G_{xxxx}$ component one should add to the Hamiltonian (\ref{Hamilt}) scalar terms linear in $B_x$ and odd in $k_x$, see the next section.

The eMCh current density is calculated in the standard way with the help of the hole distribution function $f_{\bm k}$ as follows
\begin{equation}
\label{j_general}
\delta \bm j = 2e \sum_{\bm k} \bm v(\bm k) f_{\bm k} \:,
\end{equation}
where $e>0$ is the elementary charge, the factor 2 accounts for the two valleys $H$ and $H'$, and $\bm v(\bm k)$ is the hole velocity $\hbar^{-1}\partial \varepsilon_{\bm k}/\partial \bm k$. 

In the $z$-eMChA geometry, the distribution function $f_{\bm k}$ is dependent on $k_z$ and $k^2_{\perp}$ and independent of the azimuth angle between ${\bm k}_{\perp}$ and the $x$ axis. It is helpful to change variables from $(k_z, k^2_{\perp})$ to $(k_z, \varepsilon^0_{\bm k})$ bearing in mind that
\[
k^2_{\perp} = \frac{1}{ {\cal A}_2 } \qty( \varepsilon^0_{\bm k} + \sqrt{ \Delta_2^2 + \beta^2 k^2_z } - \Delta_2 ) \:.
\]
Thus, all functions $k_z$ and $k^2_{\perp}$ are treated as dependent on $k_z$ and energy $\varepsilon^0_{\bm k}$, $\mathcal F(k_z,\varepsilon^0_{\bm k})$.
A sum of any function $\mathcal F(k_z,\varepsilon^0_{\bm k})$ over ${\bm k}$ is calculated as follows
\begin{equation}
\label{sum_to_int_kz}
\sum_{\bm k} \mathcal F(k_z,\varepsilon^0_{\bm k}) 
=g_{2D}\int\limits_{\varepsilon_m}^\infty \dd \varepsilon^0_{\bm k} \int\limits_{K_z(\varepsilon^0_{\bm k})} \dd k_z \mathcal F(k_z,\varepsilon^0_{\bm k})\:,
\end{equation}
where $g_{2D}= {1/ (8\pi^2 \mathcal A_2)}$ is the density of states for two-dimensional motion in the $(xy)$ plane. In the following we consider the case where the energy minimum $\varepsilon_m$ is small compared with the average hole energy and use the limits $[- \kappa(\varepsilon_{\bm k}),  \kappa(\varepsilon_{\bm k})] $ and $(0,\infty)$ of integration over $k_z$ and $\varepsilon^0_{\bm k}$ in equations like Eq.~(\ref{sum_to_int_kz}).

The distribution function obeys the Boltzmann kinetic equation
\begin{equation}
\label{kin_eq_general}
{e\over \hbar}\bm E \cdot \pdv{f_{\bm k}}{\bm k} + \hat{\mathcal I}^{(\rm el)}_{\bm k}\qty[f] + \hat{\mathcal I}^{(\rm inel)}_{\bm k}\qty[f] = 0 \:,
\end{equation}
where the left-hand side contains the force and collision terms respectively. The collision integral consists of two contributions describing elastic and inelastic hole scattering. 
We will solve Eq.~\eqref{kin_eq_general} by iterations up to the second order in $\bm E$ and, therefore, present $f_{\bm k}$ as a sum $f_0(\varepsilon_{\bm k})+ f_1(\bm k)+ f_2(\bm k)$ with  $f_0(\varepsilon_{\bm k})$ being the Fermi-Dirac distribution and $f_n \propto E^n$.

\section{Relaxation time approximation} 
\label{tau_approx}

We use the relaxation-time approximation taking the collision integral in the form
\begin{equation}
\hat{\mathcal I}_{\bm k}\qty[f] =  {f_{\bm k} - f_0(\varepsilon_{\bm k}) \over \tau}
\end{equation}
with $\tau$ being a constant.
This corresponds to fast energy relaxation with a rate equal to the elastic scattering rate.

Then the corrections of the first and second orders in $E_z$ are given by
\begin{eqnarray}
f_1(\bm k)&=&-e\tau E_z f_0'(\varepsilon_{\bm k}) v_z\:, \\ f_2(\bm k) &=& -{e\tau E_z  \over \hbar} \pdv{f_1(\bm k)}{k_z}\:, \nonumber
\end{eqnarray}
where $f_0'(\varepsilon) = \partial f_0(\varepsilon)/\partial \varepsilon$. Substitution of $f_{\bm k}=f_2({\bm k})$ into Eq.~\eqref{j_general} and integration by parts yields for the eMCh current density
\begin{equation}
\label{j_tau_approx}
\delta j_z = 2{e^3\qty(\tau E_z)^2  \over \hbar^3} \sum_{\bm k} f_0(\varepsilon_{\bm k}) \pdv[3]{\varepsilon_{\bm k}}{k_z}\:.
\end{equation}
An analogous result was obtained previously for 1D transport in quantum wires~\cite{Loss}. 
Expanding 
\begin{equation}
\label{f_0_expand}
f_0(\varepsilon_{\bm k}^0 + \delta\varepsilon_{\bm k}) \approx f_0(\varepsilon_{\bm k}^0)+ f_0'(\varepsilon_{\bm k}^0)\delta\varepsilon_{\bm k}\:,
\end{equation}
calculating the third derivatives of $\varepsilon_{\bm k}^0$ and $\delta\varepsilon_{\bm k}$ and integrating by parts we obtain
\begin{equation}
G_{zzzz}=  - 12  {e^3 \tau^2  \mathcal A_1 g \beta \over \hbar^3 \Delta_2} \sum_{\bm k} \eta^2(k_z) \zeta^3(k_z) f_0'(\varepsilon_{\bm k}^0) \:,
\end{equation}
where 
\begin{equation}
\zeta(k_z)={\Delta_2 \over \sqrt{\Delta_2^2+\beta^2 k_z^2}} = \sqrt{1-\eta^2(k_z)}.
\end{equation}
For degenerate hole gas with the Fermi energy $\varepsilon_F > 0$ we come to the final equation
\begin{equation}
\label{gamma_tau_approx}
G_{zzzz}
=g{\mathcal A_1\over  \mathcal A_2}{e^3 \tau^2 \over \pi^2 \hbar^3} \eta^3(\kappa_{\text F}) \:,
\end{equation}
where  $\kappa_{\text F}=\kappa(\varepsilon_{\text F})$ and $\kappa(\varepsilon)$ is defined by Eq.~\eqref{kappa_def}. 

An important result to stress is that, at small $\beta$, the eMCh current is not linear but cubic function of $\beta$. This can be understood taking $\varepsilon^0_{\bm k}$ as $\mathcal A_1 k_z^2 + A_2 k_{\perp}^2$ and the magnetic-field induced correction to the energy as $\delta \varepsilon = P k_z$, where $P = -g B_z\beta /\Delta_2$, and shifting the origin of the ${\bm k}$-space by $k^0_z = P/(2 {\cal A}_1)$. In the new frame $k'_z = k_z + k^0_z$ we obtain a fully parabolic dispersion $\varepsilon_{{\bm k}'} = \mathcal A_1 k_z^{\prime 2} + \mathcal A_2 k_{\perp}^2$ as in a centrosymmetric crystal where an eMCh current is forbidden.

A similar calculation of the $G_{xxxx}$ component with the following energy dispersion in the magnetic field $\bm B \parallel x$
\begin{equation} \label{kx}
\varepsilon_{\bm k}=\varepsilon_{\bm k}^0 +  B_xk_x(\Xi_\perp k_\perp^2 +\Xi_z k_z^2)\:,
\end{equation}
yields
\begin{equation}
G_{xxxx} = 2 {\mathcal A_1\over \mathcal A_2}{e^3\tau^2  \over \pi^2 \hbar^3} \Xi_\perp  \kappa^3_{\text F}.
\end{equation}
Since the presence of a linear-$k_x$ term is not enough to get the magnetochiral current, we took into account cubic-$\bm k$ terms in Eq.~(\ref{kx}). Note, however, that $\Xi_z$ makes no contribution to the eMCh current
because the third derivative $\partial^3(k_x k_z^2)/\partial k_x^3 =0$, see Eq.~\eqref{j_tau_approx}.

\subsection{Microscopic interpretation of eMChA}

We give here the simplest interpretation of the eMChA current~(\ref{j_tau_approx}). In the external electric field $E_z$ the equilibrium hole distribution is shifted in the ${\bm k}$-space by $\delta k_z=eE_z \tau /\hbar$. Then in the simplest description one can present the nonequilibrium distribution function as $f_0({\bm k}_{\perp}, k_z - \delta k_z)$. Let us expand this function in powers of $\delta k_z$ as follows
\[
f_0({\bm k}_{\perp}, k_z - \delta k_z) = f_0({\bm k}) - \frac{\partial f_0}{\partial k_z} \delta k_z + \frac12 \frac{\partial^2 f_0}{\partial k_z^2} (\delta k_z)^2\:.
\] 
The linear term contributes to the Ohmic current while the nonlinear contribution is
\begin{equation}
\delta j_z = e \sum\limits_{\bm k} v_z \frac{\partial^2 f_0}{\partial k_z^2} (\delta k_z)^2 =\frac{e^3 \tau^2 E_z^2}{\hbar^3} \sum\limits_{\bm k} \frac{\partial^3 \varepsilon_{\bm k}}{\partial k_z^3}  f_0(\varepsilon_{\bm k})\:.\nonumber
\end{equation}
This equation differs from Eq.~(\ref{j_tau_approx}) only by a factor of 2 which reflects the simplified character of the latter consideration.

\section{Mechanism due to elastic scattering} \label{Elastic}
Now we consider the eMCh current formed in the process of elastic scattering by short-range impurities.
In this case the collision integral reads
\begin{equation}
\label{coll_int}
\hat{\mathcal I}^{(\rm el)}_{\bm k}[f]= {2\pi\over \hbar}\mathcal N_i\sum_{\bm k'} \abs{V_{\bm k' \bm k}}^2\delta\qty(\varepsilon_{\bm k}-\varepsilon_{\bm k'}) (f_{\bm k}-f_{\bm k'}) ,
\end{equation}
where $\mathcal N_i$ is the impurity concentration, and $V_{\bm k' \bm k}$ is the matrix element of scattering by an individual impurity potential $V(\bm r)=V_0\delta(\bm r)$ given by $V_{\bm k' \bm k} = V_0 \braket{u_{k_z'}}{u_{k_z}}$,
with $u_{k_z}$ being the eigenvectors of the Hamiltonian~\eqref{Hamilt}. For the mechanism under consideration all the noneqilibrium corrections to the distribution function $f_{\bm k}$ vanish after averaging over ${\bm k}$ at the fixed energy. The role of corrections $\delta f$ dependent on the energy $\varepsilon_{\bm k}$ is analyzed in Sect.~\ref{Inelastic}.

\subsection{Inversion of the collision integral at $\bf{B=0}$} 
\label{222}

At $\bm B=0$ we obtain from Eq.~\eqref{u_eta}:
\begin{equation}
\label{V_quad}
\abs{V_{\bm k' \bm k}}^2 = {V_0^2\over 2} \qty[1+\eta(k_z)\eta(k_z')+\zeta(k_z)\zeta(k_z')].
\end{equation}
Below we use for brevity the notation $\mathcal I_{\bm k}[f]$ instead of $\mathcal I^{(\rm el)}_{\bm k}[f]$ for the elastic collision integral at ${\bm B} = 0$.

It follows from Eq.~\eqref{V_quad} that, for the short-range scattering potential, the kernel of the elastic collision integral~\eqref{coll_int} is degenerate: It is a sum of products of functions depending solely on $k_z$ or $k_z'$. This allows us to invert the operator $\hat{\mathcal I}_{\bm k}[f]$ by reducing the following  integral equation to the algebraic one
\begin{equation} \label{GI}
G(k_z, \varepsilon^0_{\bm k}) + \hat{\mathcal I}_{\bm k}[f] =0\:,
\end{equation}
where the source function $G(k_z, \varepsilon^0_{\bm k})$ satisfies the integral condition
\begin{equation} \label{norm_cond}
\sum\limits_{\bm k} G(k_z, \varepsilon^0_{\bm k}) \delta(\varepsilon^0_{\bm k} - \varepsilon) \propto \int\limits_{-\kappa(\varepsilon)}^{\kappa(\varepsilon)} \dd k_z G(k_z, \varepsilon)=0\:,
\end{equation}
which means that the number of particles of a given energy are conserved under elastic scattering.
If the source function $G(k_z, \varepsilon^0_{\bm k})$ in the kinetic equation~(\ref{GI}) does not satisfy the condition~(\ref{norm_cond})  it should be presented as a sum of the function satisfying this condition and the function $G(\varepsilon^0_{\bm k})$ dependent purely on $\varepsilon^0_{\bm k}$. In order to find the solution of the kinetic equation with the source $G(\varepsilon^0_{\bm k})$ one must replace $\hat{\mathcal I}_{\bm k}[f]$ in Eq.~(\ref{GI}) by the inelastic collision integral $\hat{\mathcal I}^{(\rm inel)}_{\bm k}[f]$, see Section~\ref{Inelastic}.
For an odd source term, $G(k_z, \varepsilon^0_{\bm k}) =-G(-k_z, \varepsilon^0_{\bm k})$, we obtain for the inverse operator
\begin{equation}
\hat{\mathcal I}_{\bm k}^{-1}\qty[G]=
{\hbar [G(k_z, \varepsilon^0_{\bm k}) + \eta(k_z) L_{G\eta}/\qty( 1-L_{\eta^2})]\over 2\pi g_{2D} \mathcal N_i V_0^2 C(k_z,\varepsilon^0_{\bm k})}\:,
\end{equation}
and for an even function $G(k_z, \varepsilon^0_{\bm k}) =G(-k_z, \varepsilon^0_{\bm k})$ we have
\begin{equation}
\hat{\mathcal I}_{\bm k}^{-1}\qty[G]=  {\hbar [G(k_z, \varepsilon^0_{\bm k}) - \zeta(k_z) L_G / L_\zeta] \over 2\pi g_{2D} \mathcal N_i V_0^2  C(k_z,\varepsilon^0_{\bm k})} \:,
\end{equation}
%\vspace{- 3 mm}
\begin{equation}
C(k_z,\varepsilon^0_{\bm k})=\kappa(\varepsilon^0_{\bm k}) + {\Delta_2\over \beta}\zeta(k_z)
\Arctanh \qty{ \eta[ \kappa (\varepsilon^0_{\bm k})]} \:,
\end{equation}
where $\Arctanh(z)=[\ln(1+z)-\ln(1-z)]/2$, and the function $L_F(\varepsilon^0_{\bm k})$  is defined for any even-$k_z$ function $F(k_z,\varepsilon^0_{\bm k})$ as follows
\begin{equation}
\label{def_L}
L_F (\varepsilon^0_{\bm k}) =\int\limits_{0}^{\kappa(\varepsilon^0_{\bm k})} \dd k_z   
{F(k_z,\varepsilon^0_{\bm k}) \over C(k_z,\varepsilon^0_{\bm k}) }\:.
\end{equation}

By using the inverse collision integral we can calculate the conductivity 
\begin{equation} \label{szz}
\sigma_{zz}=2e\sum_{\bm k}v^0_z(\bm k)\hat{\mathcal I}_{\bm k}^{-1}\qty[-ev^0_zf_0']\:,
\end{equation}
where $v_z^0 = \hbar^{-1}\partial \varepsilon_{\bm k}^0/\partial k_z$ is the hole velocity in the absence of magnetic field.
For degenerate hole statistics we have
\begin{equation}
\sigma_{zz} ={2 e^2 \hbar \over \pi \mathcal N_i V_0^2} \qty(L_{v^2} + {L_{v\eta}^2 \over 1- L_{\eta^2}})\:,
\end{equation}
where the functions $L_F$ are taken at $\varepsilon^0_{\bm k}=\varepsilon_{\rm F}$.

\subsection{Allowance for linear-$\bf B$ term in collision integral}

Scattering by impurities is affected by the magnetic field.
In the linear-$\bm B$ approximation we obtain
\begin{align}
\label{delta_I}
\delta\hat{\mathcal I}_{\bm k}\qty[f]=&
{2\pi\over \hbar}\mathcal N_i \sum_{\bm k'} \biggl[\delta\abs{V_{\bm k' \bm k}}^2 \delta\qty(\varepsilon_{\bm k}^0-\varepsilon_{\bm k'}^0) 
\\
&+\abs{V_{\bm k' \bm k}}^2\delta'\qty(\varepsilon_{\bm k}^0-\varepsilon_{\bm k'}^0) \qty(\delta\varepsilon_{\bm k}-\delta\varepsilon_{\bm k'}) \biggr] \qty(f_{\bm k}-f_{\bm k'}). \nonumber
\end{align} 
Here $\delta\varepsilon_{\bm k}$ is given by Eq.~\eqref{d_E_k}, and, since the Hamiltonian~\eqref{Hamilt} in the presence of $B_z$ is obtained from its zero-field value by the sustitution $k_z \to k_z + gB_z/\beta$, we have
\begin{equation}
\delta\abs{V_{\bm k' \bm k}}^2 = {g B_z\over \beta}\qty(\pdv{}{k_z}+\pdv{}{k_z'})\abs{V_{\bm k' \bm k}}^2,
\end{equation}
which yields
\begin{eqnarray}
&& \hspace{1 cm} \delta\abs{V_{\bm k' \bm k}}^2 =V_0^2 {g B_z\over 2\Delta_2}
\\ && \times \qty[\eta^2(k_z')-\eta^2(k_z)]\qty[\eta(k_z')\zeta(k_z)-\eta(k_z)\zeta(k_z')]\:. \nonumber
\end{eqnarray}
Passing from summation to integration over the variables $(k_z', \varepsilon_{\bm k'}^0)$ and integrating the term with $$\delta' (\varepsilon_{\bm k}^0 - \varepsilon_{{\bm k}'}^0)= - \frac{\partial \delta (\varepsilon_{\bm k}^0 - \varepsilon_{{\bm k}'}^0)}{\partial \varepsilon_{{\bm k}'}^0}$$ by parts, we get
\begin{multline}
\label{delta_I1}
\delta\hat{\mathcal I}_{\bm k}\qty[f]={2\pi\over \hbar}\mathcal N_i g_{2D} 
\\ \times \Biggl\{
\int\limits_{-\kappa(\varepsilon_{\bm k}^0)}^{\kappa(\varepsilon_{\bm k}^0)} \dd k_z' \delta\abs{V_{\bm k' \bm k}}^2  \qty[f(\varepsilon_{\bm k}^0,k_z)-f({\varepsilon_{\bm k}^0,k_z'})]
\\
+
f_{\bm k} \dv{}{\varepsilon_{\bm k}^0}\int\limits_{-\kappa(\varepsilon_{\bm k}^0)}^{\kappa(\varepsilon_{\bm k}^0)} \dd k_z' \abs{V_{\bm k' \bm k}}^2\qty(\delta\varepsilon_{\bm k}-\delta\varepsilon_{\bm k'}) 
\\  - \dv{}{\varepsilon_{\bm k}^0}\int\limits_{-\kappa(\varepsilon_{\bm k}^0)}^{\kappa(\varepsilon_{\bm k}^0)} \dd k_z' \abs{V_{\bm k' \bm k}}^2\qty(\delta\varepsilon_{\bm k}-\delta\varepsilon_{\bm k'}) f({\varepsilon_{\bm k}^0,k_z'})
\Biggr\}.
\end{multline}
Here we took into account that both $\abs{V_{\bm k' \bm k}}^2$ and $\delta\varepsilon_{\bm k}$ are independent of $\varepsilon_{\bm k}^0$ and dependent on $k_z, k_z'$ only. 

\subsection{Procedure to calculate the eMCh current}

According to Eq.~(\ref{delta_I}) or~(\ref{delta_I1}), at nonzero magnetic field the kinetic equation takes the form
\begin{equation}
\label{kin_eq_B}
{e\over \hbar}E_z\pdv{f_{\bm k}}{k_z} +\hat{\mathcal I}_{\bm k}\qty[f] +\delta\hat{\mathcal I}_{\bm k}\qty[f] = 0\:. 
\end{equation}
The equilibrium hole gas is described by the Fermi--Dirac distribution function~\eqref{f_0_expand}
satisfying the identity
\begin{equation}
\delta\hat{\mathcal I}_{\bm k}\qty[f_0(\varepsilon_{\bm k}^0)] + \hat{\mathcal I}_{\bm k}\qty[ f_0'(\varepsilon_{\bm k}^0)\delta\varepsilon_{\bm k}] =0\:,
\end{equation}
which is Eq.~(\ref{kin_eq_B}) at $E_z = 0$.

The correction to the distribution function proportional to $E_z^2B_z$ can be found by iterations of the kinetic equation~\eqref{kin_eq_B}. First of all, we find a linear-$E_z$ correction $f_{\bm k}^{(E)}$ at $B_z=0$. It is given by $f_{\bm k}^{(E)}=-eE_z\hat{\mathcal I}_{\bm k}^{-1}\qty[v_zf_0']$, see Eq.~(\ref{szz}). The required solution $\delta f_{\bm k} \propto E_z^2 B_z$  is sought as a sum of two corrections labeled $f_{\bm k}^{(E^2B)}$ and $f^{(EBE)}_{\bm k}$.

To calculate $f_{\bm k}^{(E^2B)}$ we perform the next iteration and find the correction $f_{\bm k}^{(E^2)} \propto E_z^2$ at $B_z=0$ from the equation
\begin{equation}
\label{f_E_quad}
{eE_z\over \hbar}\qty(\pdv{f_{\bm k}^{(E)}}{k_z} -\overline{\pdv{f_{\bm k}^{(E)}}{k_z}})+ \hat{\mathcal I}_{\bm k}\qty[f^{(E^2)}]=0\:.
\end{equation}
Here the bar denotes averaging over $k_z$ at a fixed energy $\varepsilon^0_{\bm k}$, namely,
\begin{equation}
\overline{F} = {1\over 2 \kappa(\varepsilon^0_{\bm k})}\int\limits_{-\kappa(\varepsilon^0_{\bm k})}^{\kappa(\varepsilon^0_{\bm k})} \dd k_z F(k_z,\varepsilon^0_{\bm k})\:.
\end{equation}
Then we include into consideration the magnetic field $B_z$ and find $f_{\bm k}^{(E^2B)}$ as a solution of the linear equation
\begin{equation}
\label{f_E_quad_B}
\delta\hat{\mathcal I}_{\bm k}\qty[f_{\bm k}^{(E^2)}] + \hat{\mathcal I}_{\bm k}\qty[f^{(E^2B)}]=0\:.
\end{equation}

In order to determine the second contribution, $f_{\bm k}^{(EBE)}$, we first find the 
correction $f_{\bm k}^{(EB)}\propto E_zB_z$. It satisfies the equation
\begin{eqnarray}
\label{kin_eq_EB}
&& \hspace{0.7 cm} {e\over \hbar}E_z\qty[ \pdv{\qty(f_0' \delta\varepsilon_{\bm k})}{k_z} -\overline{\pdv{\qty( f_0' \delta\varepsilon_{\bm k})}{k_z}}]
\\ 
&& +\ \delta\hat{\mathcal I}_{\bm k}\qty[f^{(E)}]-\overline{\delta\hat{\mathcal I}_{\bm k}\qty[f^{(E)}]}
+\hat{\mathcal I}_{\bm k}\qty[f^{(EB)}]  =0\:. \nonumber
\end{eqnarray}
Finally we substitute the correction $f_{\bm k}^{(EB)}$ to 
\begin{equation}
\label{EBE}
{eE_z\over \hbar}\pdv{f_{\bm k}^{(EB)}}{k_z} + \hat{\mathcal I}_{\bm k}\qty[f^{(EBE)}]=0
\end{equation}
and find $f_{\bm k}^{(EBE)}$.
 
It should be noted that both $f_{\bm k}^{(E)}$ and the resulting functions $f_{\bm k}^{(E^2B)}, f_{\bm k}^{(EBE)}$ are odd in $k_z$, whereas those obtained at intermediate iteration steps, $f_{\bm k}^{(E^2)}$  and $f^{(EB)}$, are even functions of $k_z$. For the mechanism due to elastic scattering all these functions satisfy the integral condition~(\ref{norm_cond}).

The eMCh current is calculated according to Eq.~\eqref{j_general} as follows
\begin{equation}
\label{d_j_3_contrib}
\delta j_z=2e\sum_{\bm k} \qty[v_z^0 \qty(f_{\bm k}^{(EBE)}+f_{\bm k}^{(E^2B)} ) + \delta v_z f_{\bm k}^{(E^2)}].
\end{equation}
Here, following Eq.~(\ref{vv0dv}) we present the hole velocity as $v_z=  v^0_z + \delta  v_z$ with
\begin{align}
&v^0_z(k_z)= \frac{1}{\hbar} \left[ 2 \mathcal A_1 k_z+ \beta\eta(k_z) \right]\:, 
\\ 
&  \delta v_z(k_z) = g B_z{ \beta \over \hbar\Delta_2}\zeta^3(k_z)\:. 
\label{d_v_B}
\end{align}

\section{The magnetochiral current in the small $\beta$ limit} 
\label{small_beta}

At small $\beta$, the  equation~\eqref{gamma_tau_approx} derived  in the relaxation-time approximation reduces to
\begin{equation}
\label{gamma_tau}
G_{zzzz} \approx g{\mathcal A_1\over  \mathcal A_2}{e^3 \tau^2 \over \pi^2 \hbar^3} z^3\:,
\end{equation}
where $z = \beta \kappa_F/\Delta_2$.

Here we go beyond the relaxation-time approximation, apply the scheme developed in the previous Section and calculate each of three contributions to the eMCh current (\ref{d_j_3_contrib}) assuming the constant $\beta$ to be small.

In the limit $\beta \to 0$, we have approximately
\begin{eqnarray}
&&C(k_z,\varepsilon^0_{\bm k}) \approx 2\kappa(\varepsilon_{\bm k}^0)\:,\:  \kappa(\varepsilon_{\bm k}^0) \approx \sqrt{\varepsilon_{\bm k}^0\over \mathcal A_1}\:,\: 
\abs{V_{\bm k' \bm k}}^2 \approx V_0^2\:, \nonumber \\ &&  \hspace{1.1 cm}\varepsilon_{\bm k}^0 \approx \mathcal A_1 k_z^2 + \mathcal A_2 k_\perp^2\:,\: \quad v_z^0 \approx {2\mathcal A_1 k_z\over\hbar}\:, 
\end{eqnarray}
and the inverted collision integral is given by
\begin{equation}
\label{tau_exact}
\hat{\mathcal I}_{\bm k}^{-1}[G] \approx -\tau(\varepsilon_{\bm k}^0) G(k_z,\varepsilon^0_{\bm k})\:,\: \tau(\varepsilon_{\bm k}^0) 
=  {2\pi \mathcal A_2 \hbar \over \mathcal N_i V_0^2 \kappa(\varepsilon_{\bm k}^0)}\:.
\end{equation}
In the magnetic-field induced correction to the energy spectrum we take into account the cubic-$\beta$ term because, as discussed above, the linear-$\beta$ correction does not result in eMChA. Therefore we take
\begin{equation}
\label{d_e_d_v}
\delta\varepsilon_{\bm k} 
\approx -{gB_z\over 2}\qty({\beta k_z\over \Delta_2})^3,
\quad
\delta v_z
\approx -{3gB_z\over 2\hbar}\qty({\beta \over \Delta_2})^3 k_z^2.
\end{equation}
The $B_z$-linear correction to the scattering matrix element squared reads
\begin{equation}
\delta\abs{V_{\bm k' \bm k}}^2 \approx {V_0^2\over 2}{g B_z\over \Delta_2}\qty({\beta k_z\over \Delta_2})^3 \qty(k_z'^2-k_z^2)\qty(k_z'-k_z).
\end{equation}
It can be neglected in the following because its contribution to the current is parametrically smaller by a factor of $\varepsilon_{\rm F}/\Delta_2 \ll1$ compared with other contributions coming from the $B_z$-linear correction~\eqref{d_e_d_v}.
As a result, only two last lines of Eq.~\eqref{delta_I1} contibute to $\delta\hat{\mathcal I}_{\bm k}\qty[f]$:
\begin{equation}
\label{delta_I_small_beta}
\delta\hat{\mathcal I}_{\bm k}\qty[f]={1\over \kappa\tau} \qty[f_{\bm k}\delta\varepsilon_{\bm k}\dv{\kappa}{\varepsilon_{\bm k}^0} + {1\over 2}\int\limits_{-\kappa(\varepsilon_{\bm k}^0)}^{\kappa(\varepsilon_{\bm k}^0)} \dd k_z' \delta\varepsilon_{\bm k'}\pdv{f({\varepsilon_{\bm k}^0,k_z'})}{\varepsilon_{\bm k}^0}].
\end{equation}

We start from calculation of the third term in the right-hand side of Eq.~(\ref{d_j_3_contrib}). The correction $f_{\bm k}^{(E^2)}$ found from Eq.~\eqref{f_E_quad} with $f_{\bm k}^{(E)} = -eE_z\tau v_zf_0'$ is given by
\begin{equation}
\label{f_E_quad_Small_beta}
f_{\bm k}^{(E^2)} 
= \qty(2\mathcal A_1{eE_z\over \hbar})^2 \tau(\tau f_0')' \qty(k_z^2 - {\kappa^2\over 3}).
\end{equation}
Substituting this function into the last term in Eq.~\eqref{d_j_3_contrib} we find its contribution to the eMChA effect
\begin{equation} \label{ve2}
G^{(v)}_{zzzz} =  - {32\mathcal A_1^2ge^3\tau\over 15\hbar^3}\qty({\beta \over \Delta_2})^3  g_{2D} \pdv{[\kappa_{\rm F}^5\tau(\varepsilon_{\rm F})]}{\varepsilon_{\rm F}}.
\end{equation}
Using the relations $\kappa_{\rm F} \propto \varepsilon_{\rm F}^{1/2}$, $\tau(\varepsilon_{\rm F}) \propto \varepsilon_{\rm F}^{-1/2}$, $\mathcal A_1\kappa_{\rm F}^2=\varepsilon_{\rm F}$,
we arive at
\begin{equation}
G^{(v)}_{zzzz} = -{8\over 15} g{\mathcal A_1\over  \mathcal A_2}{e^3 \tau^2 \over \pi^2 \hbar^3}z^3\:.
\end{equation}

Next, we search for the correction $f_{\bm k}^{(E^2B)}$. It is found from Eq.~\eqref{f_E_quad_B} to be as follows, see Eq.~\eqref{delta_I_small_beta}),
\begin{equation}
f_{\bm k}^{(E^2B)} =gB_z \qty({\beta k_z \over \Delta_2})^3  {1\over 2\kappa}  \dv{\kappa}{\varepsilon_{\bm k}^0}   f_{\bm k}^{(E^2)}\:,
\end{equation}
where $f_{\bm k}^{(E^2)}$ is given by Eq.~\eqref{f_E_quad_Small_beta}.
This allows us to calculate the second contribution in Eq.~\eqref{d_j_3_contrib}:
\begin{equation} \label{e2b}
G^{(E^2B)}_{zzzz}={16\over 105} g{\mathcal A_1\over  \mathcal A_2}{e^3 \tau^2 \over \pi^2 \hbar^3}z^3\:. 
\end{equation}

Finally we calculate the contribution related to the function $f_{\bm k}^{(EBE)}$. 
According to Eq.~\eqref{EBE} this function has the form
\begin{equation}
f_{\bm k}^{(EBE)} = -\tau {eE_z\over \hbar} \pdv{f_{\bm k}^{(EB)}}{k_z}\:.
\end{equation}
It allows us to rewrite the first contribution in Eq.~\eqref{d_j_3_contrib} as
\begin{equation} \label{tauprime}
j_z^{(EBE)}=2e^2 E_z \qty({2\mathcal A_1\over \hbar})^2\sum_{\bm k} f_{\bm k}^{(EB)}  \tau'  k_z^2\:.
\end{equation}
While deriving this equation we took into account that the function $ f_{\bm k}^{(EB)}$ satisfies Eq.~(\ref{norm_cond}).

The solution of Eq.~\eqref{kin_eq_EB} for $f_{\bm k}^{(EB)}$ reads
\begin{align}
f^{(EB)}_{\bm k} = &\frac{gB_zeE_z \tau}{2\hbar} \qty({\beta \over \Delta_2})^3 
\left[ 3 f_0' \qty(k_z^2-{\kappa^2\over 3})   \right. \\ & \left. \hspace{1 cm} +\ \mathcal A_1 \qty(2f_0''-{f_0'\over  \varepsilon_{\bm k}^0}) \qty(k_z^4-{\kappa^4\over 5}) \right]\:. \nonumber
\end{align}
Substitution of this expression to Eq.~(\ref{tauprime}) leads to 
\begin{equation} \label{ebe}
G_{zzzz}^{(EBE)} = - {2\over 105}g{\mathcal A_1\over  \mathcal A_2}{e^3 \tau^2 \over \pi^2 \hbar^3}z^3\:. 
\end{equation}
The sum of three contributions (\ref{ve2}), (\ref{e2b}) and (\ref{ebe}) yields
\begin{equation}
\label{G_elastic}
G_{zzzz} =-{2\over 5}g{\mathcal A_1\over  \mathcal A_2}{e^3 \tau^2 \over \pi^2 \hbar^3}z^3\:, 
\end{equation}
where $\tau$ is defined by Eq.~\eqref{tau_exact}. 
Comparing with the relaxation-time approximation result~\eqref{gamma_tau} we see a difference both in the sign and a factor of 2/5. 

\section{Mechanism involving inelastic scattering}
\label{Inelastic}

Now we turn to the mechanism of magnetochiral current involving the inelastic scattering. Compared to the previous section we change the attention from the asymmetric part of $\delta f_{\bm k}$ satisfying condition (\ref{norm_cond}) to the energy-dependent part $\delta f(\varepsilon_{\bm k})$ of the correction to the hole distribution function. Assuming the hole-hole collisions to be more effective than the hole energy relaxation on acoustic phonons we can describe the energy-dependent sum $f_0(\varepsilon_{\bm k}) + \delta f(\varepsilon^0_{\bm k})$ as the Fermi-Dirac distribution function $f_0(\varepsilon_{\bm k},T_h)$ characterized by the hole temperature $T_h$ different from the bath temperature $T$. Here we first briefly describe the procedure to calculate $T_h$ 
%at ${\bm B} = 0$ 
and then show how the inelastic relaxation of hole nonequilibrium distribution $f_0(\varepsilon_{\bm k},T_h)$ gives rise to an electric current proportional to $ (T_h - T) B_z$.

\subsection{Estimation of the hole effective\\ temperature $\propto E_z^2$}
The effective temperature $T_h$ can be found from the heat balance equation
\begin{equation} \label{sigmaEE}
\sigma_{zz} E_z^2 = {\cal J}\:.
\end{equation}
The left-hand side represents Joule heating produced by the passage of an electric current with $\sigma_{zz}$ being the conductivity. The right-hand side describes the energy relaxation
of the holes following acoustic-phonon scattering and has the form
\begin{equation} \label{J}
 {\cal J} =  \sum_{ {\bm k}' {\bm k} } \left( \varepsilon_{\bm k} - \varepsilon_{{\bm k}'}\right) \left( W^{({\rm ab})}_{{\bm k}', {\bm k}} -  W^{({\rm em})}_{{\bm k}, {\bm k}'}\right)\:,
\end{equation}
where $W^{({\rm ab})}_{{\bm k}', {\bm k}}$, $W^{({\rm em})}_{{\bm k}, {\bm k}'}$ are the hole scattering rates  for phonon absorption and emission processes.
Their difference is given by
\begin{align} \label{+-}
W^{({\rm ab})}_{{\bm k}', {\bm k}} -  W^{{\rm em}}_{{\bm k}, {\bm k}'} =& \frac{2 \pi}{\hbar} |M_{{\bm k}'{\bm k} }|^2   \delta(\varepsilon_{{\bm k}'}   -\varepsilon_{\bm k} -  \hbar \Omega_{\bm q}) 
\\ & \times \left[ ( f_{\bm k} - f_{{\bm k}'}) N_{\bm q} - f_{{\bm k}'} (1 - f_{\bm k}) \right] \: . \nonumber
\end{align}
Here ${\bm q} = {\bm k}' - {\bm k}$, $\Omega_{\bm q}$ and $N_{\bm q}$ are the phonon wave vector, frequency and occupation number
\begin{equation} \label{phon}
N_{\bm q} = \frac{1}{ \exp( \hbar \Omega_{\bm q}/k_B T) - 1} \:, \nonumber
\end{equation}
$M_{{\bm k}'{\bm k} }$ is the scattering matrix element.  
For the energy-dependent distribution function $f_0(\varepsilon_{\bm k},T_h)$ the term in the  brackets in Eq.~(\ref{+-}) reduces to
\begin{align}
&  \frac{{\rm e}^{(\varepsilon - \varepsilon_F)/k_B T_h} \left[ {\rm e}^{(\varepsilon' - \varepsilon)/k_B T_h}  - {\rm e}^{(\varepsilon' - \varepsilon)/k_B T}\right]}{ ({\rm e}^{(\varepsilon - \varepsilon_F)/k_B T_h} + 1)({\rm e}^{(\varepsilon - \varepsilon')/k_B T_h} + 1) ({\rm e}^{(\varepsilon' - \varepsilon)/k_B T} - 1)}  \nonumber \\ &\approx - \frac{T_h - T}{T}  \frac{\hbar \Omega_{\bm q}/k_B T}{ {\rm e}^{\hbar \Omega_{\bm q}/k_B T} - 1} f_0 (\varepsilon) \left[ 1 - f_0(\varepsilon') \right] \:, \nonumber
\end{align}
where $\varepsilon = \varepsilon^0_{\bm k}$, $\varepsilon' =\varepsilon^0_{{\bm k}'}$. 

For the degenerate statistics, $\varepsilon_F \gg k_B T$,
a reasonable estimation of ${\cal J}$ in Eqs.~(\ref{sigmaEE}), (\ref{J}) is
\begin{equation} \label{right0}
{\cal J} \sim \Delta \varepsilon \frac{ T_h - T}{T} \frac{\rho(\varepsilon_F) k_B T}{\tau_{in}}
=  k_B (T_h - T) \frac{\rho(\varepsilon_F) \Delta \varepsilon}{\tau_{in}}\:, 
\end{equation}
where $\Delta \varepsilon = \min{(\hbar \Omega_{k_F}, k_B T)}$, and $\rho(\varepsilon)$ is the 3D density of states. The characteristic inelastic-scattering time $\tau_{in}$ is defined by
\begin{equation} \label{acphon}
\frac{1}{\tau_{in}} = \frac{2 \pi}{\hbar} \sum_{{\bm k'}} |M_{{\bm k}'{\bm k}}|^2 \delta( \varepsilon^0_{{\bm k}'}- \varepsilon^0_{\bm k})
\end{equation} 
for $\varepsilon^0_{\bm k} = \varepsilon_F$. Equations (\ref{sigmaEE}) and (\ref{right0}) allow one to estimate the heating of the hole gas.

{\subsection{Current driven by energy relaxation}
The energy-dependent nonequilibrium function $f(\varepsilon_{\bm k},T_h)$ makes no contribution to the current (\ref{j_general}). However, an electric current appears due to the inelastic relaxation of this distribution to $f(\varepsilon_{\bm k},T) \equiv f_0(\varepsilon_{\bm k})$. The current is given by
\begin{equation} \label{jzas}
\delta j_z = - e \sum_{\bm k}\tau v^{(0)}_z(k_z) {\cal I}^{(ne)}_{\bm k}\{ f\}\:,
\end{equation}
where  the inelastic collision integral has the form
\begin{eqnarray} \label{Ine2}
&&{\cal I}^{(ne)}_{\bm k}\{ f\} = \frac{2 \pi}{\hbar} \sum_{ {\bm k}' }  |M_{{\bm k}' {\bm k} }|^2\\ && \left\{ \left[  (f_{\bm k} - f_{{\bm k}'}) N_{\bm q} +  f_{\bm k} (1 - f_{{\bm k}'}) \right] \delta(\varepsilon_{{\bm k}'}- \varepsilon_{\bm k} + \hbar \Omega_{\bm q}) \right. \nonumber\\ &&  \left.  + \left[ ( f_{\bm k}  - f_{{\bm k}'}) N_{\bm q} -  f_{{\bm k}'} (1 - f_{\bm k})  \right]\delta(\varepsilon_{{\bm k}'} - \varepsilon_{\bm k} - \hbar \Omega_{\bm q}) \right\} \:, \nonumber
\end{eqnarray}
with $f_{\bm k} = f_0(\varepsilon_{\bm k},T_h)$ and $\varepsilon_{\bm k} = \varepsilon^0_{\bm k} + \delta \varepsilon_{\bm k}$, see Eqs.~(\ref{e0}) and (\ref{d_E_k}). It is clear that the current is contributed by the odd-in-$k_z$ part of ${\cal I}^{(ne)}_{\bm k}\{ f\}$. For simplicity we used the relaxation time approximation for deriving the antisymmetric component of the hole distribution function $f^{(2)}_{\bm k}$ and get $f^{(2)}_{\bm k}=-\tau \mathcal I^{(ne)}_{\bm k}\{ f\}$.

Substituting the collision integral into Eq.~(\ref{jzas}) we can reduce this equation to
\begin{align} \label{jzas2}
&\delta j_z = - \frac{2 \pi e \tau}{\hbar}  \sum_{{\bm k}, {\bm k}'}  |M_{{\bm k}' {\bm k} }|^2 \left[ v^{(0)}_z(k_z) - v^{(0)}_z(k'_z) \right]  \\ & \times \left[ ( f_{\bm k}  - f_{{\bm k}'}) N_{\bm q} -  f_{{\bm k}'} (1 - f_{\bm k})  \right]\delta (\varepsilon_{{\bm k}'} - \varepsilon_{\bm k} - \hbar \Omega_{\bm q})  \:. \nonumber
\end{align}

For an estimation of the current magnitude we simplify in the collision integral the dispersion (\ref{e0}) to $\varepsilon^0_{\bm k} = {\cal A} k^2$ and take into account only the cubic term in the expansion 
of $\delta \varepsilon(k_z)$, see Eq.~\eqref{d_e_d_v}. Then the expressions in the sums (\ref{J}) and (\ref{jzas2}) differ by the multipliers $(\varepsilon_{\bm k} - \varepsilon_{{\bm k}'})$ and 
\begin{eqnarray}
&& {g B_z\over \varepsilon_F} \left[ v^{(0)}_z(k_z) \left( \frac{\beta k_z}{\Delta_2}\right)^3 - v^{(0)}_z(k'_z)  \left( \frac{\beta k'_z}{\Delta_2}\right)^3 \right] \nonumber \\ && \hspace{1.5 cm} \sim {g B_z\over \varepsilon_F} \left( \frac{\beta}{\Delta_2}\right)^3 \frac{k^2}{\hbar} \left( \varepsilon_{\bm k} - \varepsilon_{{\bm k}'} \right) \:.\nonumber
\end{eqnarray}
It follows then that the current (\ref{jzas2}) can be estimated as 
\[
\delta j_z \sim e \tau  {g B_z\over \varepsilon_F} \left( \frac{\beta}{\Delta_2} \right)^3 \frac{ \kappa_F^2}{\hbar} {\cal J}\:.
\]
For the simplified energy dispersion the conductivity reads
\[
\sigma_{zz} \sim \frac{e^2 \tau \varepsilon_F \kappa_F}{\hbar^2}
\]
and we finally obtain
\begin{equation} \label{jzas3}
\delta j_z \sim g \frac{e^3 \tau^2}{\hbar^3}  \left( \frac{\beta \kappa_F}{\Delta_2}\right)^3 E_z^2 B_z\:.
\end{equation}
One can see that the obtained estimation of the magnetochiral current for the second mechanism has the same order as the contribution~\eqref{G_elastic}.

\section{Discussion} 

\label{Disc_concl}We begin the discussion with a general symmetry analysis of the eMChA effect studied in this paper. Tellurium is a crystal with chiral (or enantiomorphic) structure. By definition, a chiral periodic solid (or molecule) is non-superimposable with its mirror image and has a ``handedness''. Two modifications of a chiral structure that are mirror-like to each other are called enantiomorphic. In tellurium crystals, the two mirror modifications are characterized by the space groups $D_3^4~(P3_{1}21)$ and $D_3^6~(P3_{2}21)$. Among 32 crystallographic point groups, 11 are enantiomorphic, namely, ${\cal F} = C_1, C_2, D_2, C_4, D_4, C_3, D_3$ (quartz, tellurium), $C_6, D_6, T$ and $O$.

In this regard, the question arises which of the coefficients $G^{(n)}$ in Eqs.~(\ref{Macro}) coincide and which differ in sign for the two enantiomorphs. To answer this question, consider the achiral point group $D_{3h}$, which differs from the $D_3$ group by the presence of a symmetry plane $\sigma_h$ and includes 12 operations $g \in D_{3h}$. The $D_{3h}$ symmetry allows nonzero terms in (\ref{Macro}) with coefficients $G^{(3)}, G^{(8)}$ and $G ^{(10)}$. Consequently, these three coefficients describe an electric current nonlinear in ${\bm E}$ and linear in ${\bm B}$ with its sign independent of the enantiomorphic modification. The remaining seven coefficients $G^{(n)}$ describe magnetochiral currents with opposite directions for the $D_3^4$ and $D_3^6$ phases. This way of separating the chiral and achiral contributions to the electric current is applicable for ten enantiomorphic crystal classes ${\cal F}$, except for the $O$ class. Each of them can be associated with an achiral point group ${\cal F}_a \ni {\cal F}$, which has no spatial inversion center and which admits nonzero coefficients $G_{ijkl}$ in Eq.~(\ref{ji}). These coefficients describe achiral transport, whereas the additional coefficients arising in the ${\cal F}$ group are chiral. Chiral and achiral nature of the coefficients can be readily determined from the behavior of physical quantities in the left and right parts of Eqs.~(\ref{Macro}) under reflection in the $\sigma_h$ plane. Indeed, under this operation the component $\delta j_z$ changes sign, but the product $E_z^2 B_z$ is invariant which means that the coefficient $G^{(1)}$ is chiral. At the same time, the product $E_x^2 B_y$ changes sign upon reflection $\sigma_h$, as does the component $\delta j_x$. Therefore, the coefficient $G^{(3)}$ describes achiral transport.

Let us list the achiral groups corresponding to the above ten chiral groups: ${\cal F}_a = C_s, C_{2v}, D_{2d}, C_{4v}, D_{4d}, C_{3v}, D_{3h}, C_ {6v}, D_{6d}$ and $T_d$. As another example, consider the chiral group $T$ (sillenite Bi$_{12}$SiO$_{20}$, bismuth germanate Bi$_{12}$GeO$_{20}$) and the corresponding achiral group $T_d$. The $T_d$ symmetry allows for the current $\delta j_x$ terms proportional to $(E_y^2 - E_z^2)B_x$ and $(E_y B_y - E_z B_z) E_x$. In addition, the $T$ group has chiral contributions proportional to $|{\bm E}|^2 B_x, ({\bm E} \cdot {\bm B}) E_x$ and $E_x^2 B_x$. The symmetry point transformation which can be used to divide between chiral and achiral coefficients is the reflection in the plane $\sigma_v \parallel (110)$.

As for the enantiomorphic group $O$, it has no partner ${\cal F}_a \ni O$ without an inversion center. Adding a reflection plane to the group $O$ leads to the  $O_h$ group in which all the coefficients $G_{ijkl}$ are equal to zero. Hence, for the $O$ group, all the coefficients $G_{ijkl}$ in the expansion (\ref{ji}) are chiral.

Note that the BiTeI crystal has the achiral trigonal symmetry $C_{3v}$ and allows a nonreciprocal rectification effect $\delta j_x \propto E_x^2 B_y$ \cite{BiTeBr}, which however is not a magnetochiral effect.

In Sections \ref{tau_approx}--\ref{Inelastic}, we have considered successively various models and mechanisms of the eMChA effect: the approximation of a constant relaxation time, the general procedure for calculating the magnetochiral current for different elastic and inelastic  relaxation times, and the approximation of a small chiral parameter $\beta$. A derivation of the exact expression for the current beyond the fixed relaxation time approximation cannot be obtained analytically and is outside the scope of this work. However, the carried-out study shows that the magneto-chiral current $\delta j_z$ in tellurium for a degenerate hole gas can be described by 
$\delta j_z = G_{zzzz}E_z^2 B_z$ with
\begin{equation} \label{general_eq}
G_{zzzz}= c g \frac{{\cal A}_1}{{\cal A}_2} \frac{e^3 \tau^2}{\pi^2 \hbar^3} \left( \frac {\beta \kappa_F}{\Delta_2} \right)^3 \:,
\end{equation}
where $c$ is a factor of the order of unity. This means that the resistance of tellurium $R$ has the nonreciprocal chiral contribution
\begin{equation}
R = R_0(1+\gamma j_z B_z), \qquad \gamma = {G_{zzzz}\over \sigma^2},
\end{equation}
where $R_0$ is the resistance in the absence of magnetic field, and $\sigma$ is the conductivity. For an estimation we ignore the difference between $\mathcal A_1$ and $\mathcal A_2$. 
Taking the conductivity as $\sigma = pe^2\tau/m^*$ where $m^*=2\mathcal A_1/\hbar^2$, the coefficient $g$ in Eq.~\eqref{Hamilt} as $g = g^* \mu_{\rm B}$, where $\mu_{\rm B}$ is the Bohr magneton and $g^*$ is the effective $g$-factor, and noting that the Fermi wavevector is related to the hole concentration as $\kappa_{\rm F}^3 = 3\pi^2 p$, we get
\begin{equation}
\gamma\approx {3 \mu_{\rm B} g^*{m^*}^2\over  e p} \qty({\beta \over\hbar \Delta_2})^3.
\end{equation}
We use the parameters suitable for Te: $\beta = 2.5 \times 10^{-8}$~{eV cm}, $\Delta_2 = 63$~{meV}, $m^*=0.2m_0$, $g^*=1$, and the hole concentration $p=10^{16}$~cm$^{-3}$. Then we obtain that the ratio $\beta \kappa_{\rm F}/\Delta_2 \approx 0.27 \ll 1$, and one can apply the approximate equation~\eqref{general_eq} for the estimation. The result yields 3$\times 10^{-7}$~cm$^2$\,T$^{-1}$A$^{-1}$ for the magneto-induced rectification coefficient $\gamma$.

The eMCh current measurements presented in Figs.~3 and~4 in Ref.~\cite{Rikken2019} were performed in the following two geometries: (i)~the electric current measured in the $x$ direction at ${\bm E} \parallel x$ and the magnetic field vector lying in the $(xy)$ plane,  (ii) ${\bm j}, {\bm E} \parallel z$, and the magnetic field in the $(xz)$ plane. It follows from the general equations (\ref{Macro}) that, in these two setups, one has
\begin{align} 
& \label{jxtheta} \delta j_x =  G_{xxxx} E_x^2 B \sin{\theta_y} = ( G^{(5)} + G^{(7)} ) E_x^2 B \sin{\theta_y}\:,  
\\ & \label{jztheta}  \delta j_z = G_{zzzz} E_z^2 B \sin{\theta_x} =  G^{(1)} E_z^2 B \sin{\theta_x} \:, 
\end{align}
where $B = \abs{\bm B}$, $\theta_y$ is the angle between the vector ${\bm B}$ lying in the $(xy)$ plane and the $y$ axis, $\theta_x$ is the angle between the vector ${\bm B}$ lying in the $(xz)$ plane and the $x$ axis.

According to Rikken and Avarvari~\cite{Rikken2019} their measurements on tellurium show that $3\gamma_{xxxx} \approx \gamma_{xxxy}$ and $12 \gamma_{zzzx} \approx \gamma_{xxxy}$, and $\gamma_{zzzz} \ll \gamma_{zzzx}$. These results are in complete contradiction to the phenomenological equations (\ref{Macro}), (\ref{jxtheta}) and (\ref{jztheta}) derived for  D$_3$ symmetry crystals. Indeed, the symmetry predicts that $\gamma_{xxxy} = \gamma_{zzzx} = 0$ while
the component $\gamma_{zzzz}$ is allowed. This is a key difficulty in comparing the derived theory with the experiment \cite{Rikken2019} and an additional experimental work is needed on the study of the chiral transport in tellurium crystals.

In Ref.~\cite{Rikken2019}, a theoretical estimate of the $\gamma$ value is also given. The equation for $\gamma$ is derived in the framework of a model where the linear in $k_zB_z$ term in the hole energy dispersion is taken into account only. As stressed in Section \ref{tau_approx} this term does not lead to the eMChA and one needs to include the higher-order term $k_z^3 B_z$ in the hole Hamiltonian, as unambiguously follows from Eq.~(\ref{j_tau_approx}) for $\delta j_z$. Moreover, 
the $k_zB_z$-linear term $\delta \varepsilon_{\bm k} = \chi k_z B_z$ in the hole dispersion is given by $\chi= - g\beta/\Delta_2$ with an estimate for tellurium $\abs{\chi} = 3.7 \times 10^{-32}$~{J m/T}.
The value of $\abs{\chi}$ assumed in Ref.~\cite{Rikken2019} is $\sim 40$ times larger.

So far, we have examined the effect of a static electric field ${\bm E}$. It is easiest to generalize the theory to the case of a time-dependent field $E_z(t) = E^{(0)}_z \cos{\omega t}$ in the constant-time approximation for frequencies satisfying the condition $\omega \ll \varepsilon_F/\hbar$, while the product $\omega \tau$ may be arbitrary. To find the distribution function $f_{\bm k}(t)$, the derivative $\partial f_{\bm k}/ \partial t$ must be added to the left side of the kinetic equation (\ref{kin_eq_general}). Omitting calculations, we present the result. The formula (\ref{j_tau_approx}) for $\omega \neq 0$ becomes
\begin{equation} \label{jzomega}
\delta j_z (\omega) = \frac{\delta j_z (0)}{1 + \omega^2 \tau^2}\:,
\end{equation}
where $\delta j_z (0)$ is the magnetochiral current in a static electric field. The alternating electric field induces not only a dc current (\ref{jzomega}), but also a current at double frequency $2 \omega$. For the second harmonic generation we have
\begin{equation} \label{2omega}
j_z^{2 \omega}(t) = j_{z; 2 \omega} {\rm e}^{- 2{\rm i} \omega t} +  j_{z; -2 \omega} {\rm e}^{2{\rm i} \omega t}\:,
\end{equation} 
where the complex amplitude is given by
\[
 j_{z;2 \omega} =  j^*_{z; -2 \omega} = \frac12 \frac{\delta j_z (0)}{(1 - {\rm i} \omega \tau)(1 - 2{\rm i} \omega \tau)}\:.
\]
It should be mentioned that experimentally it is convenient to detect the magneto-chiral current by measuring the amplitude of the second harmonic at $\omega \tau \ll 1$ \cite{Rikken2019}.

In fact, Eq.~(\ref{jzomega}) describes the phenomenon is called magneto-photogalvanic effect (MPGE). In general, it is described by the following phenomenological equation \cite{Ivchenko1988,Belkov2005,IvchenkoBook,Weber_2008,Baltz}
\begin{equation} \label{lincirc}
j_i = G_{ijkl} (\omega) \{ E_j E^*_k \} B_l +  G^{(\rm circ)}_{klm} R_l B_m \:,
\end{equation}
where ${\bm E}$ is the complex amplitude of the radiation electric field and 
\[
\{ E_j E^*_k \} =  \frac12 (E_j E_k^* + E^*_j E_k) \:, \quad
{\bm R}={\rm i} ({\bm E} \times {\bm E}^*)\:.
\]
The first and second contributions in the right-hand side of Eq.~\eqref{lincirc} represent the so-called linear and circular MPGE. At zero frequency (static electric field) the coefficients $G_{ijkl} (\omega)$ coincide with the coefficients $G_{ijkl}$ in Eq.~(\ref{ji}). The electric field of the electromagnetic wave is complex and the magneto-photogalvanic current contains an additional contribution described by the pseudotensor ${\bm G}^{(\rm circ)}$ if the circular polarization of the exciting light is nonzero. The circular MPGE has first been observed in the achiral GaAs crystal~\cite{Andrianov}. Similarly to the dc effect~(\ref{Macro}) the coefficients $G_{ijkl} (\omega)$ and ${\bm G}^{(\rm circ)}$ can be divided into chiral and achiral ones.
Let us consider the products $R_i B_j$ which transform in $D_{3h}$ according to $(A'_2 + E'') \times ( A'_2 + E'') = A'_1 + 2 E'' + (A'_1 + A'_2 + E')$:
\begin{eqnarray}
&&R_z B_z~ (A'_1); \quad R_xB_x + R_y B_y ~ (A'_1); \\ && R_x B_y - R_y B_x~(A'_2); \nonumber \\ &&  R_x B_x - R_y B_y, - R_x B_y - R_y B_x ~(E'); \nonumber\\&&  R_z B_y, - R_z B_x~(E'');  \quad R_y B_z, - R_x B_z~(E'')\:.\nonumber
\end{eqnarray}
Thus, an achiral contribution to the current is given by 
\begin{align}
&\delta j_x = G_1^{({\rm circ})} ( R_x B_x - R_y B_y) \:, \nonumber \\
&\delta j_y = - G_1^{({\rm circ})}  (R_x B_y + R_y B_x) \:. \nonumber 
\end{align}
%\[
%\delta j_x = G_1^{({\rm circ})} ( R_x B_x - R_y B_y) \:,\]
%\[\delta j_y = - G_1^{({\rm circ})}  (R_x B_y + R_y B_x) \:.
%\]
In the $D_3$ symmetry, additional chiral terms appear
\begin{eqnarray}
&& \delta j_x = G_2^{({\rm circ})} R_zB_y + G_3^{({\rm circ})} R_yB_z\:,\\ && \delta j_y = - G_2^{({\rm circ})}R_z B_x - G_3^{({\rm circ})} R_x B_z \:, \nonumber\\ && \delta j_z = G_4^{({\rm circ})} (R_x B_y - R_y B_x)\:. \nonumber
\end{eqnarray}

It is instructive to describe the hierarchical sequence of point-group categories: among 21 crystal classes lacking inversion symmetry, 18 are gyrotropic and, as mentioned above, 11 are enantiomorphic. 
All noncentrosymmetric crystals allow nonzero coefficients $G_{ijkl}$ in Eq.~(\ref{ji}) and $G_{ijkl} (\omega),  G^{(\rm circ)}_{klm}$ in Eq.~(\ref{lincirc}). We remind that the gyrotropic classes allow nonzero components of the rank 3 tensors $\gamma_{ijk}$ antisymmetric under exchange of one pair of its indices or, equivalently, the rank 2 pseudotensors. In the gyrotropic crystals, there exist coefficients in Eq.~(\ref{lincirc}) that relate the current vector components with pseudovector combinations of the products of $E_j E^*_k B_l$ and describe the magneto-gyrotropic photogalvanic effects \cite{Belkov2005,Weber_2008,Tarasenko2007}. And finally, in the chiral crystals there are coefficients which have different signs for the different enantiomorphic modifications. Recently the magneto-chiral photogalvanic current $\bm j \propto \bm B \times \bm R$ has been studied in bulk tellurium~\cite{Exp_Regens} in both terahertz and infrared ranges at indirect intraband and direct intersubband optical transitions in the valence band, respectively.
\vspace{2 mm}

\section{Summary} 
\label{Summary}

We have derived the theory of eMChA effect in tellurium which shows an intricate combination of chirality and magnetism. Macroscopic phenomenological relationship is established between the electric current density and  products of the magnetic field and bilinear combinations of the electric field strength. Two microscopic mechanisms of the effect are considered, one with allowance for elastic scattering processes only and the other where the eMChA current is formed in the course of hole gas heating and its energy relaxation. In the purely elastic mechanism,  the general formalism is developed to calculate the eMChA current at arbitrary ratio between the camel-back dispersion parameter $\beta$, Fermi energy and valence-band splitting $2 \Delta_2$. 

The exact result is obtained in the limit of small $\beta$. It shows the same order of magnitude of the magneto-induced rectification coefficient $\gamma$ as that obtained in the simple relaxation-time approximation; however, the value and even the sign of $\gamma$ are different.

An attention is attracted to the difference between the achiral and chiral contributions to the magneto-induced rectification which, respectively, coincide and are opposite in sign in the two enantiomorphic modifications of chiral crystals. 

Relationship between the eMChA and magneto-induced photogalvanic effects is discussed and the chiral and achiral coefficients describing these effects in tellurium are identified. The developed theory of eMChA is compared with the available experimental data.

\acknowledgements
L.~E.~G. acknowledges support from the Deutsche Forschungsgemeinschaft (DFG, German Research Foundation) Project No. Ga501/19-1.
E.~L.~I. acknowledges support from the Russian Science Foundation (Project No. 22-12-00211).

\bibliography{EMChA_bib}

\end{document}